\documentclass[11pt]{article}

\usepackage[normalem]{ulem}
\usepackage{amsmath, amssymb,amsthm,bm,bbm,fullpage}
\usepackage{mathrsfs}
\usepackage{cite}
\usepackage{colortbl}
\usepackage[all]{xy}
\usepackage{epsfig}
\usepackage{subfigure}
\usepackage{graphics}
\usepackage{multirow}
\usepackage{lineno}
\usepackage{setspace}
\usepackage{caption}

\usepackage{graphicx}

\theoremstyle{remark} 
	
	\doublespace

\begin{document}

\begin{centering}
{\huge
\textbf{Eco-evolutionary tradeoffs in the dynamics of prion strain competition}
}
\bigskip
\\
Saul Acevedo$^{1}$  and  Alexander J. Stewart$^{2*}$
\\
\bigskip
\end{centering}
\begin{flushleft}
{\footnotesize
$^1$ Department of Biology, University of Houston, Houston, TX, USA
\\
$^2$ School of Mathematics and Statistics, University of St Andrews, St Andrews, KY16 9SS, United Kingdom
\\
$^*$ E-mail: ajs50@st-andrews.ac.uk
}
\end{flushleft}
\textbf{Prion and prion-like molecules are a type of self replicating aggregate protein that have been implicated in a variety of neurodegenerative diseases. Over recent decades the molecular dynamics of prions have been characterized both empirically and through mathematical models, providing insights into the epidemiology of prion diseases, and the impact of prions on the evolution of cellular processes. At the same time, a variety of evidence indicates that prions are themselves capable of a form of evolution, in which changes to their structure that impact their rate of growth or fragmentation are replicated, making such changes subject to natural selection. Here we study the role of such selection in shaping the characteristics of prions under the nucleated polymerization model (NPM). We show that fragmentation rates evolve to an evolutionary stable value which balances rapid reproduction of $PrP^{Sc}$ aggregates with the need to produce stable polymers. We further show that this evolved fragmentation rate differs in general from the rate that optimizes transmission between cells. We find that under the NPM, prions that are both evolutionary stable and optimized for transmission have a characteristic length of $3n$, i.e three times the critical length below which they become unstable. Finally we study the dynamics of inter-cellular competition between strains, and show that the eco-evolutionary tradeoff between intra- and inter-cellular competition favors coexistence.} 
\\
\\
\section*{Introduction}
Prions are a well known class of aggregative proteins responsible for a number of neurodegenerative disorders, including Creutzfeldt-Jakob disease (CJD) in humans, bovine spongiform encephalopathy (BSE) in cattle and scrapie in sheep. Illnesses is caused by misfolding of the PrP protein ($PrP^{Sc}$), which coerces the benign conformation ($PrP^{c}$) into an abnormal state. The misfolded $PrP^{Sc}$ then aggregates within the brain, forming transmissible spongiform encephalopathies (TSEs) \cite{scheckel2018prions} which impair brain function and ultimately result in death. As a result, the biochemistry of prion replication has been studied thoroughly, and the molecular dynamics have even been characterized mathematically via the nucleated polymerization model (NPM) \cite{nowak1998prion, masel1999quantifying, masel2001measured, payne1998spatial, tanaka2006physical, lemarre2019generalizing}. 

The role of natural selection in shaping the biochemistry of prion strains has become increasingly clear over recent decades \cite{li2010darwinian, krishnan2005structural, collinge2007general, collinge2010prion}. Strains accumulate ``mutations'' in their conformation which alter their rate of reproduction, e.g. by changing the rate at which they co-opt $PrP^{c}$ into their aggregate, or the rate at which they fragment into multiple aggregates sharing the biochemical properties of the parent. The result is replication coupled with heritable variation, which provides the basis for a form of Darwinian evolution in which the protein aggregates themselves are the basic unit of replication. The question that immediately arises is: What kind of prions will this natural selection produce?

We answer this question by adapting the NPM to study the evolutionary tradeoffs faced by prion strains. We characterize the dynamics of inter-and intracellular competition between strains, and show that, in a well-mixed environment, selection acts to produce strains that most efficiently exploit the $PrP^{c}$ resource produced by a cell. Next we show that when strains compete to invade new cells, selection produces an entirely different optimum, since strains must minimize their probability of extinction when rare. And so prion strains face a tradeoff between their ability to infect new cells, and their ability to efficiently make use of the $PrP^{c}$ resources within a cell. 

In order to understand the consequences of this tradeoff, we study the the ecological dynamics of competing prion strains under a compartmental model of cell infection. We show that competition between strains that are optimized to infect new cells or to co-opt $PrP^{c}$ with in a cell respectively leads to coexistence, and in some circumstances to sustained oscillations with repeated waves of infection and reinfection. Our results show that com petition between prion strains can generate complex dynamics, with potential consequences for the epidemiology of prion diseases, and for the eco-evolutionary dynamics of the amyloid world more generally.

\section*{Results}

We study models of both intra- and inter-cellular prion dynamics. First, we employ the well-studied nucleated polymerization model (NPM)~\cite{nowak1998prion, masel1999quantifying, masel2001measured} to explore the intra-cellular evolutionary dynamics of prions. The NPM assumes that two forms of the prion protein exist: the monomeric $PrP^{c}$ conformation and the pathogenic aggregate form $PrP^{Sc}$. Aggregates grow linearly, with existing aggregates converting monomers and incorporating them into the aggregate structure.  Replication of an aggregate occurs when it fragments into two separate daughter aggregates. The resulting daughter aggregates inherit the fragmentation rate of their mother, however we also assume that spontaneous ``mutations'' can occur that alter the fragmentation rate of a single prion (Figure 1). 

\begin{figure} \centering
\includegraphics[scale=0.33]{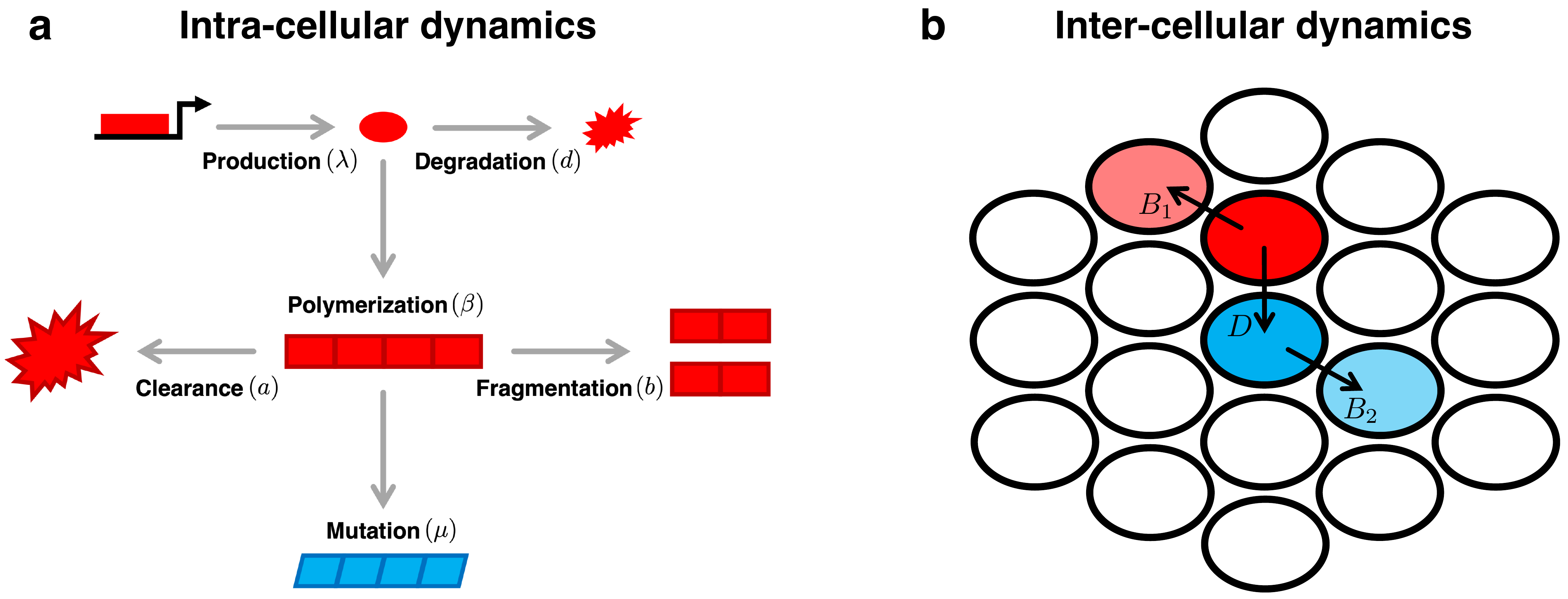}
\caption{\small\textbf{Intra- and inter-cellular prion dynamics:} a) Within a cell we adapt the nucleated polymerization model (NPM) to describe the evolution of prion strains. We model the molecular dynamics of prions as a Markov process with transitions in which aggregates undergo the processes of polymerization at rate $\beta$, fragmentation at rate $b$, degradation at rate $d$n and clearance at rate $a$. We also assume that mutations can occur, at rate $\mu$ in which the fragmentation rate $b_k$ of a polymer $k$ is perturbed by an amount $\Delta$ compared to its parent. This represents the emergence of a new strain (blue) from a resident strain (red). b) We model the dynamics of inter-cellular competition via a compartmental model in which two strains $P_1$ (red) and $P_2$ (blue) invade uninfected cells (white) at rates $B_1$ and $B_2$ respectively. We also assume that the first strain can invade the second strain, which occurs at rate $D$.}
\end{figure}

We analyse the evolutionary dynamics of the NPM, and compare our results to a stochastic version of the model, simulated using the Gillespie algorithm.  Under the stochastic model, we assume $X$ is the number of monomeric $PrP^{c}$ and $Y_i$ is the number of $PrP^{Sc}$ aggregates of size $i$ (Figure 1). Monomers are produced by the cell ($X \rightarrow X+1$) at rate $\lambda$ and degraded ($X \rightarrow X-1$) at rate $d$. Similarly, a $PrP^{Sc}$ aggregate is extended by one monomer unit ($Y_i \rightarrow Y_{i-1}$, $Y_{i+1} \rightarrow Y_{i+1}+1$) with rate $\beta$. Growth of the aggregate by assimilation of a monomer results in $X \rightarrow X-1$. Each aggregate of size $i$ has $i-1$ links connecting its $PrP^{c}$ subunits together, and each of them is assumed to fragment with fixed rate $b$. Therefore, an aggregate of size $i$ fragments at rate $b(i-1)$ into two smaller daughter aggregates of length $j$ and $i-j$. If the length of either of the daughters is below a critical size $n$, e.g. if one of the daughters has $j<n$, that prion disintegrates into $j$ monomers. We assume that, following fragmentation, conformation changes analogous to mutation can occur which alter the biochemical properties of the prion. In particular we assume that changes in the fragmentation rate of a daughter aggregate occur with rate $\mu$. Mutations are assumed to have small effects, i.e. a mutated daughter is assumed to have a new fragmentation rate that is a perturbation about the aggregation rate of the mother. Thus different prions can have different fragmentation rates. Finally, cellular degradation of individual aggregates ($Y_i \rightarrow Y_{i-1}$) occurs at rate $a$.The model parameters are summarized in Table 1. The ODE formulation of the NPM is given in the SI.

\begin{table}[ht]
\caption{Intra- and Inter-cellular Model parameters}
\centering
\begin{tabular}{l l l l}
\hline
Parameter & Definition & Modal \\ [0.5ex] 
\hline
$a$ & Clearance rate of polymers & NPM \\
$\beta$ &  Monomer to polymer conversion rate & NPM  \\
$b$& Fragmentation rate & NPM \\
$\lambda$& Monomer production rate & NPM\\
$d$ & Monomer degradation rate& NPM   \\ 
$n$ & Minimum polymer length (critical size) & NPM  \\ 
\hline
\hline
$B_1$ & Rate of strain 1 infecting uninfected cells  & Compartmental  \\ 
$B_2$ & Rate of strain 2 infecting uninfected cells  & Compartmental \\ 
$D$ & Rate of strain 1 infecting strain 2 cells  & Compartmental  \\ 
$\nu_1$ & Death rate of strain 1 infected cells  & Compartmental  \\ 
$\nu_2$ & Death rate of strain 2 infected cells  & Compartmental  \\ 
\hline
\end{tabular}
\label{table:nonlin}
\end{table}

We then study a compartmental model of inter-cellular competition between prion strains, developed based on our analysis of the NPM. Under this model two prion strains $P_1$ and $P_2$ compete to infect cells. Under this model the strains invade and infect previously uninfected cells at rate $B_1$ and $B_2$ respectively, while strain 1 is able to infect strain 2 at rate $D$ (Figure 1). Finally we assume that infected cells die at rate $\nu_1$ and $\nu_2$ respectively.

\subsection*{Prion evolution}

We begin by studying evolution of the prion traits $(b,a,\beta)$ which correspond to the fragmentation rate, the clearance rate and the growth rate of $PrP^{Sc}$ aggregates within a cell. Evolution of these traits is said to occur if perturbations to the structure of a prion, which alter their value, is passed on to its daughters following fragmentation. Since such a change would alter the birth and/or death rate of the daughter prions, such a change corresponds to heritable variation and provides the basis for a form of evolution. We can analyse the evolutionary dynamics of these traits by adopting an approach similar to adaptive dynamics \cite{Mullon:2016aa}, in which we assess the growth rate of a rare new strain which is perturbed from the resident strain by a small amount.

\begin{figure} \centering
\includegraphics[scale=0.5]{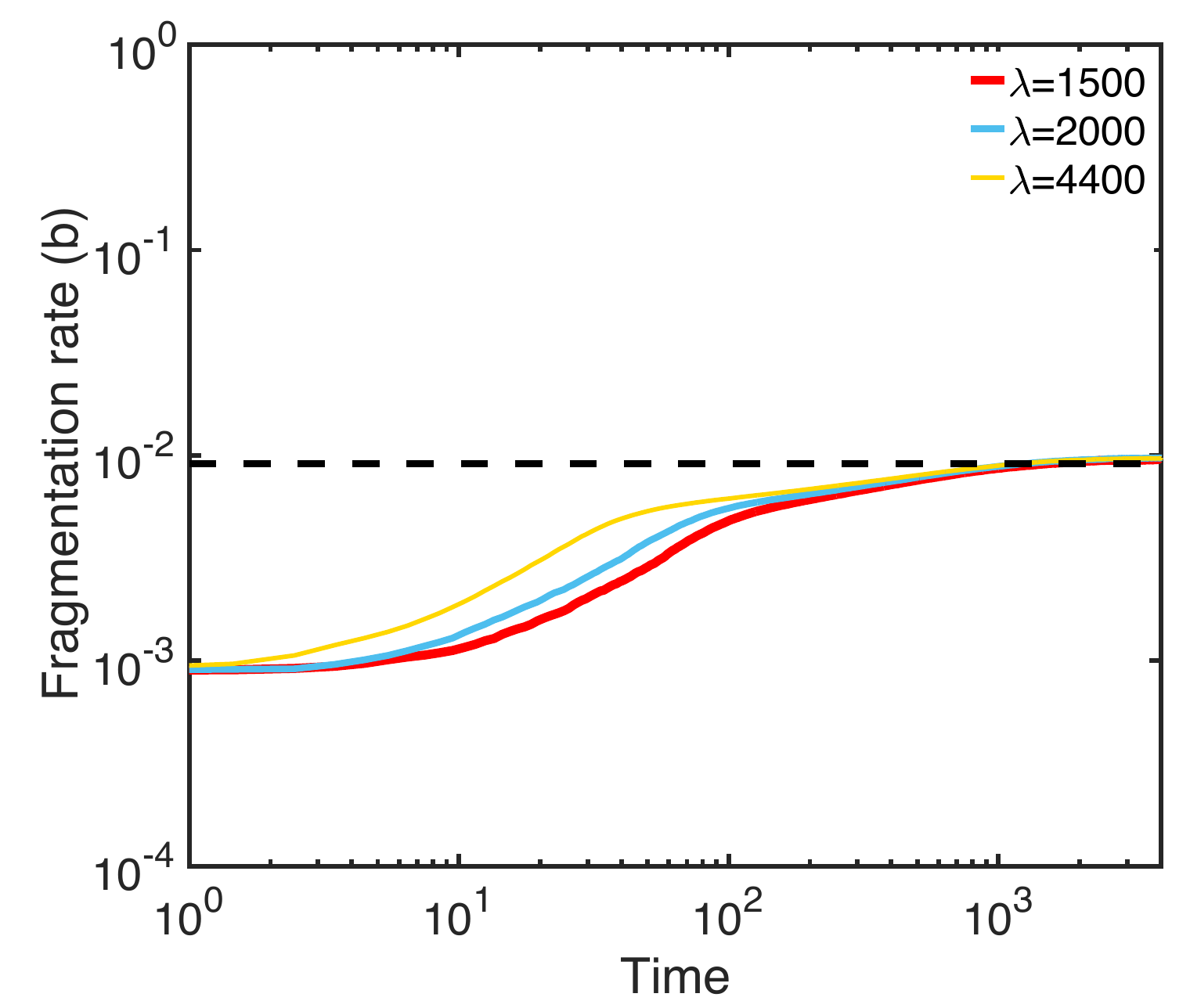}
\caption{\small\textbf{The evolution of fragmentation rate:} Prion fragmentation rate evolves to the evolutionary optimum. We simulated the stochastic NPM using the Gillespie algorithm. When evolution of fragmentation rate  $b$ is allowed the system reaches the evolutionary optimum $b_{opt}$ for different values of $\lambda$. Parameters used are $n$ = 6, $a$ = 0.05, $\beta$ = 0.015, $d$ = 4. Simulation data is an average of 1000 simulations.}
\end{figure}

We show (see SI Section 1) that under the NPM, any mutant strain which decreases the equilibrium amount of free monomer $PrP^{c}$ is able to invade and replace a resident strain. And so the evolutionary dynamics of prions lead to optimal use of the resource $PrP^{c}$. Considering the  parameters $(b,a,\beta)$, individually, we find that increasing accumulation rate $\beta$ and decreasing clearance rate $a$ will always evolve (since they produce larger, longer lived prions and so reduce the overall death rate of the strain). And so these parameters are constrained to their physical maximum and minimum respectively by the process of natural selection. The fragmentation rate $b$, in contrast, has an evolutionary optimum. Since more rapid fragmentation increases the rate of prion reproduction (if the resulting fragments exceed the critical length $n$) but also the rate of prion death (since if prions are too short when they fragment, both offspring may be shorter than the critical length $n$, resulting in both being removed from the population), there is a tradeoff. We find that the evolutionary stable fragmentation rate $b_{opt}$ is given by

\begin{equation}
b_{opt}=\frac{a}{\sqrt{n(n-1)}}
\end{equation}
\\
which depends only on the prion clearance rate $a$ and the critical length $n$. We simulated prion evolution using the stochastic NPM described above (Figure 2) and show that strains do indeed evolve to the optimum fragmentation rate Eq. 2, in a manner that does not depend on the rate of monomer production.

Having shown that natural selection acts to maximize exploitation of $PrP^{c}$ by $PrP^{Sc}$, we explored the kinds of prion that this process would produce (Figure 3). Naively we might expect natural selection to maximize the number of individual $PrP^{Sc}$ aggregates, however we show that in general this is not the case (see SI Section 1).

\begin{figure} \centering
\includegraphics[scale=0.34]{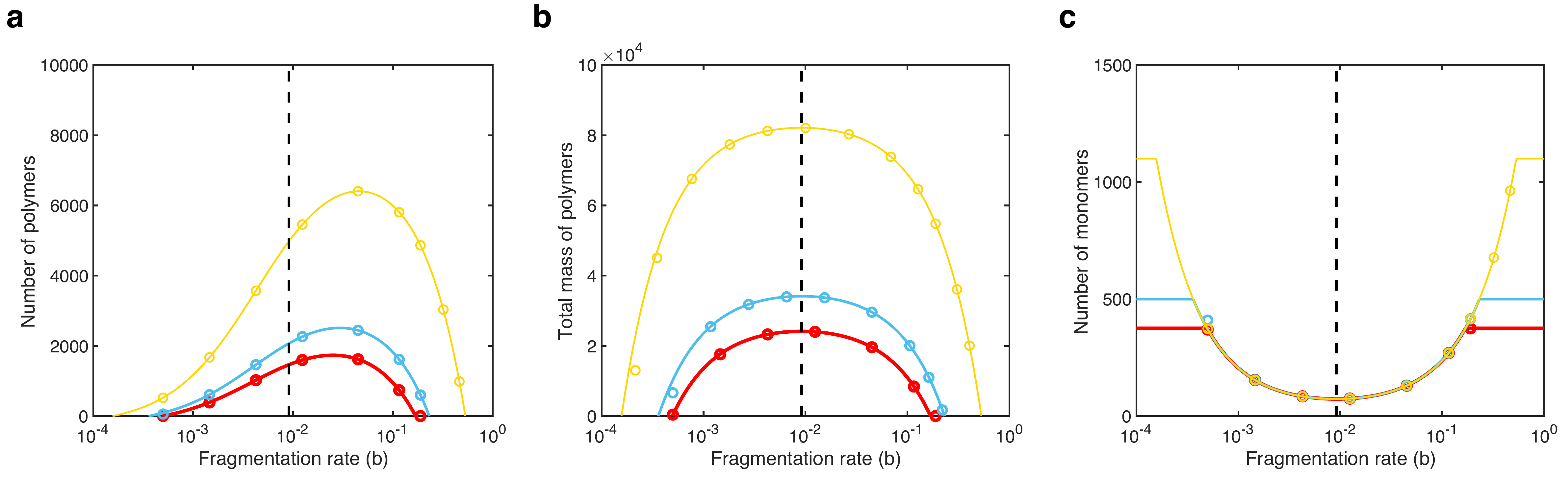}
\caption{\small\textbf{Prion abundance:}  (a) The fragmentation rate that maximizes equilibrium number of $PrP^{Sc}$ aggregates for both simulations (points) and the deterministic NPM equilibrium (lines) does not coincide with the evolutionary stable fragmentation rate (vertical dashed line) (b) The fragmentation rate that maximizes equilibrium number of $PrP^{Sc}$ subunits contained in aggregates coincides with the evolutionary stable fragmentation rate as predicted (d) Similarly fragmentation rate that minimizes equilibrium number of $PrP^{c}$ monomer coincides with the evolutionary stable fragmentation rate as predicted. We simulated the stochastic NPM using the Gillespie algorithm. Parameters are as indicated i Figure 2. Simulation data is an average of 1000 simulations..}
\end{figure}

Figure 3 shows how the number of aggregates, the total mass of aggregates, and the number of free monomers vary with fragmentation rate $b$. We see that while the optimum fragmentation rate does not maximize the number of $PrP^{Sc}$  aggregates, it simultaneously maximizes the mass the total mass of protein contained in $PrP^{Sc}$ while minimizing the number of free monomers (see also SI Section 1). And so we can view natural selection as maximizing the total number of individual $PrP^{Sc}$ proteins.

\subsection*{Optimizing transmission}

Our results on the evolution of prion traits have so far been constrained to competition between strains within a cell. However, as with any epidemiological phenomenon, we must also consider the dynamics of transmission. We study the transmission of prions between cells by considering the probability of invasion of a single $PrP^{Sc}$ aggregate into a perviously uninfected cell (see SI Section 2). We show that the optimal rate of \emph{transmission}, $b^\dagger$ can be approximated by

\begin{equation}
b^\dagger\approx \frac{\beta\lambda}{d}\frac{1}{(2n-1)^2}
\end{equation}
\\
Eq. 2 and Eq. 1 are not the same in general (Figure 4), and indeed we show in the SI (Section 2) that their ratio is approximated by

\begin{equation}
\frac{b^\dagger}{b_{opt}}\approx\frac{\beta\lambda}{4n da}
\end{equation}.
\\
Since intra-cellular natural selection tends to maximize $\beta$ while minimizing $a$, we might expect these respective optima to differ significantly, resulting in slow prion transmission when evolution acts on timescales that lead to optimization for intra-cellular competition. And so prions face an evolutionary tradeoff analogous to the virulence-transmission tradeoff faced by many pathogens \cite{bonhoeffer1996curse, bonhoeffer1994mutation}. 

\begin{figure} \centering
\includegraphics[scale=0.5]{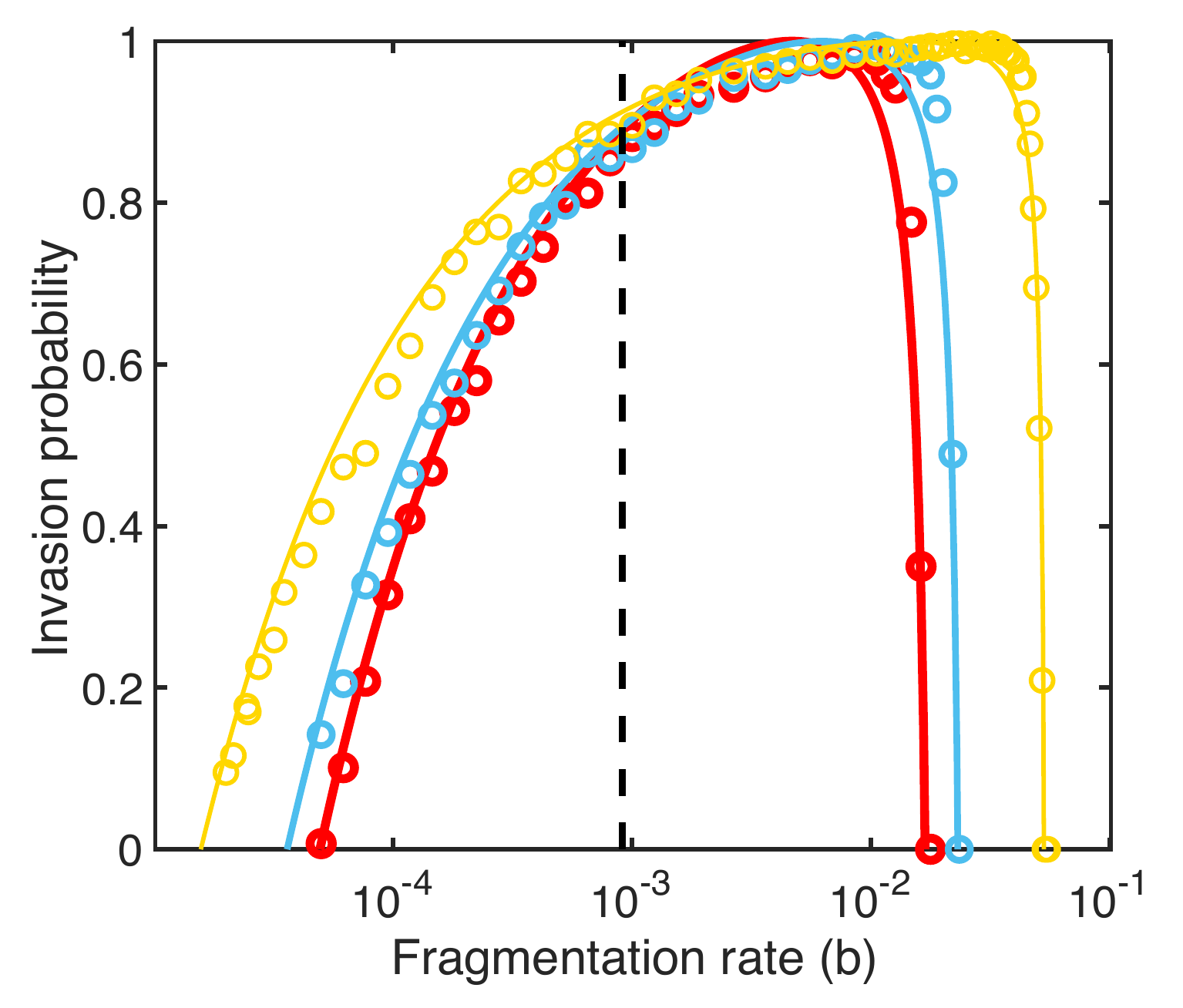}
\caption{\small\textbf{Invasion of uninfected cells:} We simulated the stochastic NPM using the Gillespie algorithm, initializing each simulation with a single $PrP^{Sc}$  aggregate drawn from the equilibrium distribution (points) and compared the results to the analytical invasion probability (see SI Section 2).  We see that the fragmentation rate that maximizes the invasion probability differs from the evolutionary stable fragmentation rate (black dashed line). Parameters used are $n$ = 6, $a$ = 0.05, $\beta$ = 0.015, $d$ = 4. Simulation data is an average of 1000 simulations.}
\end{figure}

As we have argued above, the other prion traits $a$ and $\beta$ evolve to their minimum and maximum respectively, and so are determined by physical constraints. Eq. 4 allows us to conclude that the most successful prion strains are those  for which $\frac{\beta\lambda}{4n da}\approx 1$, and so cannot be invaded by intra-cellular mutations but also maximize their rate of transmission. The characteristics of such a strain can be determined by calculating the average aggregate size $\bar{s}$. Such prions, which simultaneously optimize for inter- and intra-cellular competition, have average length $\bar{s}\approx 3n$ (see SI Section 2).

\subsection*{Competition between strains}

Having shown that there is a tradeoff between transmissibility and intra-cellular competition, we next consider the consequence for strain competition at the cellular level. We developed a compartmental model in which a class of uninfected cells $S$ are characterized by logistic growth, and two prion strains $P_1$ and $P_2$ can invade the cells. We consider a scenario in which a highly transmissible strain $P_2$ is invaded by a mutant $P_1$ that is less transmissible, but able to displace $P_2$ via intra-cellular competition. The resulting compartmental model is as follows

\begin{eqnarray}
\nonumber \frac{dS}{dt}&=&S(1-S)-S(B_1P_1+B_2P_2)\\
\nonumber \frac{dP_1}{dt}&=&B_1SP_1+P_IP_2D-\nu_1P_1\\
\frac{dP_2}{dt}&=&B_2SP_2-P_IP_2D-\nu_2P_2
\end{eqnarray}
\\
where the parameters are as described in Table 1. This model permits stable coexistence of stains (Figure 5) either at a constant abundance or as limit cycles (Figure 5). Limit cycles require that the dynamics of uninfected cells are much faster that the dynamics of prion infection (see SI Section 3).

\begin{figure} \centering
\includegraphics[scale=0.4]{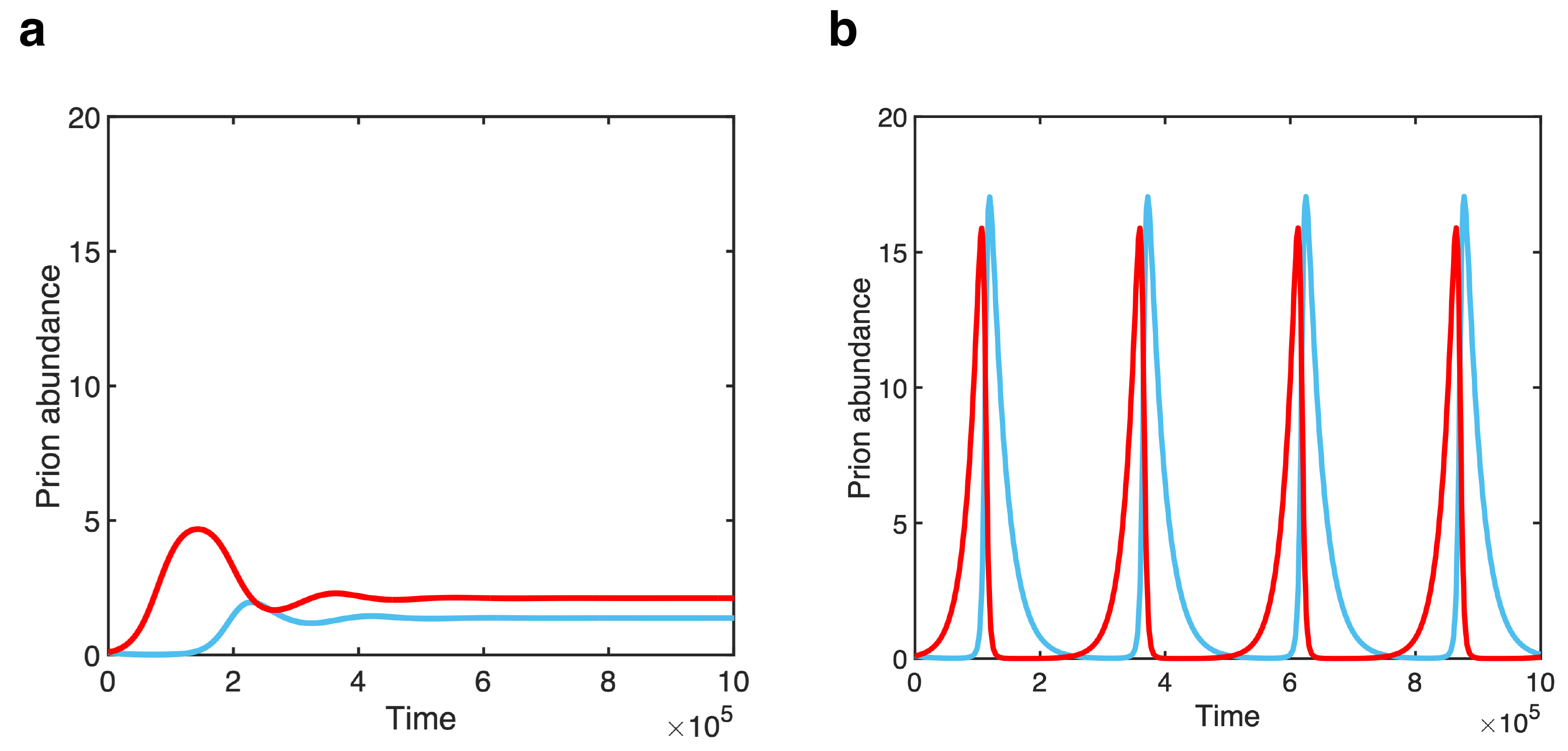}
\caption{\small\textbf{Inter-cellular competition} a) The compartmental model of intra-cellular competition (Eq. 4) produces stable coexistence when strain 1 (blue) can invade strain 2 (red) but strain 2 is better at invading uninfected cells than strain one, which occurs when the first strain has the evolutionary stable fragmentation rate $b_{opt}$ while the second has the optimal fragmentation rate for transmission $b^\dagger$. b) When uninfected cell dynamics are much faster than prion dynamics (see SI) the same process produces stable limit cycles, in which strain 2 invades, only to be replaced by strain 2, before both die out, in a pattern similar to Lotka-Volterra dynamics. Parameters used are $B_1=0.01$, $B_2=0.1$, $\nu_1=\nu_2=0.05$ and $D=0.02$.}
\end{figure}

As expected, we also observe coexistence in a spatial simulation of the same model (see Figure S1).

\section*{Discussion}

Prion and prion-like molecules have been studied for decades \cite{prusiner1998prions}, both because of their role in neurodegenerative disease and, more recently for their role in shaping the evolutionary dynamics of the organisms they infect \cite{rutherford1998hsp90, masel2013q, nelson2018evolutionary} and as the target of natural selection themselves \cite{li2010darwinian, krishnan2005structural, collinge2007general, collinge2010prion}. Our results show that natural selection will act, in a well-mixed, intra-cellular environment, to optimize the exploitation of $PrP^{c}$ by $PrP^{Sc}$. However we also show that the ability to successfully infect other cells corresponds to a different optimum fragmentation rate, and so prions face an eco-evolutionary tradeoff analogous to the virulence-transmissibility tradeoff faced by many pathogens \cite{bonhoeffer1996curse, bonhoeffer1994mutation}. We then show that inter-cellular competition between strains leads to coexistence, consistent with empirical observations or prion epidemiology \cite{ledur}.

Our results hold for the nucleated polymerization model (NPM) of prion dynamics, and as such provide a simplified and idealized description of a more complex molecular process. We show that even under such a simplified model, natural selection produces complex and counter-intuitive results. We predict that the tradeoff between transmissibility and intra-cellular competition results in the most successful prion strains having a characteristic size of roughly three times their critical length $n$. Although it is not clear that evolution will necessarily produce prions with these characteristics, our  analysis predicts that where the combination of natural selection and physical constraints produce such molecules, they will be both stable and able to spread.
Our compartmental model of inter-cellular prion dynamics captures the tradeoff between transmissibility and intra-cellular competition at the cellular level, and shows that it leads to coexistence between strains. This is in contrast to strain coexistence in a well-mixed environment (i.e within a cell) which is not possible under the NPM (see also SI Section 1.2) although such coexistence can be seen in modified versions of the model such as the template assistance model \cite{lemarre2019generalizing}. The compartmental model of inter-cellular prion dynamics also permits limit cycles, provided that infection dynamics are slow compared to the dynamics of uninfected cell birth and death. While such a scenario is unrealistic in the context of brain cells (where prion diseases have been identified) it may be relevant in the broader context of amyloids.

Prions are a subclass of amyloids, a group of aggregative proteins that are characterized by a fibrillar, $\beta$-sheet rich structure~\cite{aguzzi2009transcellular, aguzzi2016cell, scheckel2018prions}. Many amyloids are not pathogenic and provide essential functions \cite{maji2009functional, otzen2019functional}. However amyloids also comprise other  prion-like proteins called prionoids, responsible for a number of neurodegenerative diseases \cite{aguzzi2009transcellular, aguzzi2016cell, scheckel2018prions}, such as amyloid-$\beta$ which has been implicated in the development of Alzheimer's disease \cite{finder2007amyloid, roychaudhuri2009amyloid, busche2020synergy}. The key distinction between prions and prionoids is that prionoids are not transmissible between individuals. And so understanding what determines the transmissibility of an amyloid has important practical implications. More generally amyloids are understood as a source of primitive molecular replicators~\cite{maury2009self, maury2015origin, maury2018amyloid}. While it is generally accepted that RNA was the first self-replicating molecule, the high instability of RNA would have likely rendered it inactive in the extreme environment of the prebiotic soup~\cite{bernhardt2012rna}. In contrast, prions are quite stable and can persist in extreme environments, even at high temperatures~\cite{jung2003prions}. And so understanding the process of prion evolution could shed light on how self-replicating molecules could give rise to the first life.

\clearpage

\section*{Supporting Information}

\section*{Competition between prion strains}

We first look at the competition between prion strains in a well-mixed environment, e.g. within a given cell. We then look at the conditions for invasion of an empty cell, and finally at the inter-cellular dynamics under a compartmental model.

\subsection*{Selection gradient for an invading strain}

We derive the selection gradient faced by a novel strain invading a resident strain at equilibrium. The selection gradient describes the growth rate for a mutant of small effect, and is used to characterize evolution under the framework of adaptive dynamics when the resident population is very large, i.e when $\bar{y}\gg 1$.
To assess the selection gradient, we derive the invasion conditions for novel prion strains under the ODE model of Masel et. al \cite{masel1999quantifying}, under which the monomer abundance $x$, the total abundance of polymers $y$ and the total abundance of polymerized sub-units, $z$ change according to

\begin{eqnarray}
\nonumber \frac{dx}{dt}&=&\lambda-dx-\beta xy+n(n-1)by\\
\nonumber \frac{dy}{dt}&=&-ay+bz-(2n-1)by\\
\frac{dz}{dt}&=&\beta xy-az-n(n-1)by
\end{eqnarray}
\\
We consider a resident prion strain $(a,\beta,b)$ at equilibrium such that

\begin{equation}
\bar{x}=\frac{a^2 + a b(2n-1) + b^2 n(n-1)}{b \beta}
\end{equation}
\\
and we consider a mutant strain $(a^*,\beta^*,b^*)$ introduced into a cell at low frequency. In order to assess the ability of the mutant to invade we look at the behavior of the prion abundance $(y^*,z^*)$ when $x^*=\bar{x}$. The mutant strain evolves according to

\begin{eqnarray}
\nonumber \frac{dy^*}{dt}&=&-a^*y^*+b^*z^*-(2n-1)b^*y^*\\
\frac{dz^*}{dt}&=&\beta^* \bar{x}y^*-a^*z^*-n(n-1)b^*y^*
\end{eqnarray}
\\
The mutant strain will invade if the equilibrium $(0,0)$ is unstable. The eigenvalues $\Lambda$ of the system at $(0,0)$ are the solution to

\begin{equation}
(a^*+(2n-1)b^*+\Lambda)(a^*+\Lambda)-b^*(\beta^*\bar{x}-n(n-1)b^*)=0
\end{equation}
\\
i.e.

\begin{equation}
\Lambda=\frac{2a^*+(2n-1)\pm \left((2a^*+(2n-1)b^*)^2-4a^*(a^*+(2n-1)b^*)+4b^*(\beta^*\bar{x}-n(n-1)b^*)\right)^{1/2}}{2}
\end{equation}
\\
and the lead eigenvalue is positive iff

\begin{equation}
a^*(a^*+(2n-1)b^*)-b^*(\beta^*\bar{x}-n(n-1)b^*)<0
\end{equation}
\\
note that the abundance of monomers for the new strain at equilibrium is

\begin{equation}
\bar{x}^*=\frac{(a^*)^2 + a^* b^*(2n-1) + (b^*)^2 n(n-1)}{b^* \beta^*}
\end{equation}
\\
and so the condition for invasion is

\begin{equation}
b^*\beta^*\bar{x}^*-b^*\beta^*\bar{x}<0
\end{equation}
\\
If mutations are local such that the mutant strain is a small perturbation $\delta$ around the resident value then the condition for invasion for an arbitrary trait $s_i$ is

\begin{equation}
\frac{\partial \bar{x}}{\partial s_i}<0
\end{equation}
\\
and equilibrium occurs when the monomer population is minimized i.e.

\begin{equation}
\frac{\partial \bar{x}}{\partial s_i}=0
\end{equation}
\\
which is equivalent to a zero selection gradient for the monomer frequency. If we take the case of the fragmentation rate $b$, differentiating Eq. 2 we find an evolutionary stable rate when

\begin{equation}
b_{\text{opt}}=\frac{a}{\sqrt{n(n-1)}}
\end{equation}
\\
which is the result given in the main text. This equilibrium can be seen to be stable if we consider a resident $b=b^*+\delta$ and a mutant $b^*$, then from Eq. 5 the mutant will invade when rare. Similarly the evolutionary stable value for the polymerization rate is $\beta^*\to\infty$ while the evolutionary stable polymer degradation rate is $a=0$, neither of which are physically realistic -- which suggests that, if $a$ and $\beta$ are able to evolve, they will reach their physically allowable minimum and maximum respectively.

Finally we note that the equilibrium prion abundance $\bar{z}$ is give by

\begin{eqnarray}
\nonumber \bar{z}&=&\frac{b\beta\lambda-b^2dn(n-1)-abd(2n-1)-a^2d}{ab\beta}
\end{eqnarray}

which can be written as

\[
\bar{z}=\frac{\lambda}{a}-\frac{d}{a}\bar{x}
\]
\\
and so when $b$ is such that $\bar{x}$ is minimized, $\bar{z}$ is maximized.

\subsection*{Two-strain dynamics}

Next we describe the conditions for two strains to coexist. We consider a resident $(a,\beta,b)$ and a mutant $(a^*,\beta^*,b^*)$, and extend the model of Masel et. al \cite{masel1999quantifying} to account for two competing strains  

\begin{eqnarray}
\nonumber \frac{dx}{dt}&=&\lambda-dx- x(\beta y+\beta^* y^*)+n(n-1)(by+b^*y^*)\\
\nonumber \frac{dy}{dt}&=&-ay+bz-(2n-1)by\\
\nonumber \frac{dy^*}{dt}&=&-a^*y^*+b^*z^*-(2n-1)b^*y^*\\
\nonumber \frac{dz}{dt}&=&\beta xy-az-n(n-1)by\\
 \frac{dz^*}{dt}&=&\beta^* xy^*-a^*z^*-n(n-1)b^*y^*
\end{eqnarray}
\\
Solving Eq. 12 we find only three possible solutions. Either $\bar{x}=\lambda/d$, $\bar{y}=\bar{y}^*=\bar{z}=\bar{z}^*=0$. Or else $\bar{z}=\bar{y}=0$ and 

\begin{eqnarray}
\nonumber\bar{x}^*&=&\frac{(a^*)^2 + a^* b^*(2n-1) + (b^*)^2 n(n-1)}{b^* \beta^*}\\
\nonumber\bar{y}^*&=&\frac{b^*(\beta^*\lambda-b^*dn(n-1))+a^*b^*d(1-2n)-(a^*)^2d}{a^*\beta^*(a^*+b^*(2n-1))}\\
\bar{z}^*&=&\frac{b^*(\beta^*\lambda-b^*dn(n-1))+a^*b^*d(1-2n)-(a^*)^2d}{(a^*b^*\beta^*}
\end{eqnarray}
\\
or else $\bar{z}^*=\bar{y}^*=0$ and 

\begin{eqnarray}
\nonumber\bar{x}&=&\frac{a^2 + a b(2n-1) + b^2 n(n-1)}{b \beta}\\
\nonumber\bar{y}&=&\frac{b(\beta\lambda-bdn(n-1))+abd(1-2n)-a^2d}{a\beta(a+b(2n-1))}\\
\bar{z}&=&\frac{b(\beta\lambda-bdn(n-1))+abd(1-2n)-a^2d}{ab\beta}
\end{eqnarray}
\\i.e. coexistence is not possible. If we finally look at the eigenvalues associated with the second equilibrium we recover the characteristic polynomial

\begin{eqnarray}
\nonumber\left[-((a + \Lambda) (a + b (2 n-1)+\Lambda) (d + \bar{y} \beta+\Lambda)) +   -b (d + \Lambda) (b (n -1) n - \bar{x} \beta)\right]\times\\
\left[(a^*+(2n-1)b^*+\Lambda)(a^*+\Lambda)-b^*(\beta^*\bar{x}-n(n-1)b^*)\right]=0
\end{eqnarray}
\\
Here the first bracket is the characteristic polynomial for the eigenvalues of the 3-D system Eq. 1 and the second bracket is the characteristic polynomial Eq. 4 for a rare invader. And so we find that the invasion conditions for a strain at a stable equilibrium are once again determined by the  gradient $\frac{\partial \bar{x}}{\partial s_i}$.

\section*{Invasion of an empty cell}

Next we consider the conditions for a single rare prion strain to invade an empty cell. We first discuss the stability of the equilibrium $\bar{x}=\lambda/d$, $\bar{y}=\bar{z}=0$. We then consider the extinction probability for a branching process associated with the stochastic model described in the main text. 

\subsection*{Necessary condition for invasion}

In order to determine the conditions for invasion we assess the stability of the equilibrium $\bar{x}=\lambda/d$, $\bar{y}=\bar{z}=0$ for Eq. 1. The characteristic polynomial for the eigenvalues associated with this equilibrium is

\begin{equation}
(d + \Lambda) (b^2 (n - 1) n - b\lambda \beta/d - (a + \Lambda) (a + \Lambda + b (2n-1 ))) =0
\end{equation}
\\
which has solution $\Lambda=-d$ and

\begin{equation}
\Lambda=\frac{-(2 a + b  ( 2 n-1)) \pm\left(b(b+ 4\lambda \beta/d)\right)^{1/2}}{2}
\end{equation}
\\
which has a positive solution iff

\begin{equation}
b^2 n( n-1)  +b(   a  ( 2 n-1)- \lambda \beta/d)+ a^2<0
\end{equation}
\\
and so the breakage rate $b$ must lie in the range

\begin{equation}
-\gamma-\left(\gamma^2-b_{\text{opt}}^2\right)^{1/2}<b<-\gamma+\left(\gamma^2- b_{\text{opt}}^2\right)^{1/2}
\end{equation}
\\
where

\begin{equation}
\gamma=\frac{a  ( 2 n-1)- \lambda \beta/d}{2n( n-1)}
\end{equation}
\\
The only viable values of $b$ occur if $\gamma<0$ i.e. if

\begin{equation}
a <\frac{\lambda \beta}{d ( 2 n-1)}
\end{equation}
\\
and so Eq 19-21 give the conditions for a  prion strain to successfully invade an empty cell.

\subsection*{Optimal invasion rate}

Next we consider the extinction probability under a stochastic model of prion dynamics, for a population founded by a single individual. For simplicity we assume that each prion in the population has the average length

\begin{equation}
\bar{s}= n-\frac{1}{2}+\sqrt{\frac{1}{4}+\frac{\beta \lambda}{bd}}
\end{equation}
\\
derived in Masel et. al. \cite{masel1999quantifying}. We assume that each prion gan give birth to either 0, 1 or 2 offspring. Wirth probability $a$ the prion degrades and produces 0 offspring. With probability $b(\bar{s}-2n+1)$ a prion brakes into two viable fragments and produces two offspring. With probability $2b(n-1)$ a prion breaks into one viable and one inviable fragment, and produces one offspring. The probability of extinction $\eta$ under the branching process describing the system is given by the solution to

\begin{equation}
\bar{p}\eta=a+2b(n-1)\eta +b(\bar{s}-2n+1)\eta^2
\end{equation}
\\
which is

\begin{equation}
\eta=\frac{a+b(\bar{s}-2n+1)}{2b(\bar{s}-2n+1)}+\sqrt{\left(\frac{a+b(\bar{s}-2n+1)}{2b(\bar{s}-2n+1)}\right)^2-\frac{a}{b(\bar{s}-2n+1)}}
\end{equation}
\\
where

\begin{equation}
\bar{p}=a+2b(n-1)+b(\bar{s}-2n+1)
\end{equation}
\\
normalizes the transition probabilities. Eq. 25 simplifies to give an invasion probability $p_I=1-\eta$

\begin{equation}
p_I=1-\frac{a}{b(\bar{s}-2n+1)}
\end{equation}
\\
The relationship between  invasion probability and breakage rate is shown in main text Figure 4.

\clearpage

We can now determine the optimal invasion rate by finding the value of $b$ that maximizes Eq 26. Differentiating with respect to $b$ gives

\begin{equation}
\frac{p_I}{db}=\frac{a}{b^2(\bar{s}-2n+1)}\left(1-\frac{\beta\lambda}{2bd}\frac{1}{(\bar{s}-2n+1)\sqrt{\frac{1}{4}+\frac{\beta \lambda}{bd}}}\right)
\end{equation}
\\
and the value of $b$ which maximizes the invasion probability is given by

\begin{equation}
b^\dagger=\frac{\beta\lambda}{d}\left[-2+\frac{2n-1}{\sqrt{n(n-1)}}\right]
\end{equation}
\\
If we follow Masel et. al. \cite{masel1999quantifying} and make the approximation

\begin{equation}
\bar{s}\approx n-\frac{1}{2}+\sqrt{\frac{\beta \lambda}{bd}}
\end{equation}
\\
We recover the solution

\begin{equation}
b^\dagger=\frac{\beta\lambda}{d(2n-1)^2}
\end{equation}
\\
which is the expression given in the main text.

\subsection*{Tradeoff between invasion and competition}

Eq 28 and Eq 11 give different optimal values competition and invasion. And so there is a tradeoff between a successfully strategy for intra-cellular and inter-cellular competition. Indeed we characterize the extent of this tradeoff by defining $\alpha=\frac{b\dagger}{b_{opt}}$. And so our expression for $\alpha$ is

\begin{equation}
\alpha=\frac{\beta\lambda}{da}\left(2n-1-2\sqrt{n(n-1)}\right)
\end{equation}
\\
and we see that misalignment between inter- and intra-cellular optimization is greatest when prions grow rapidly ($\beta$ is large) and die slowly ($a$ is small). Finally note that Taylor expansion gives

\[
2n-1-2\sqrt{n(n-1)}=\frac{1}{4n}+O(n^{-2})
\]
\\
and so we can approximate

\begin{equation}
\alpha\approx\frac{\beta\lambda}{4nda}
\end{equation}
\\
which is the expression given in the main text.
\subsection*{Optimal prion parameters}

A prion that evolves to $b_{opt}$ such that physical constraints of $n$, $\beta$ and $a$ produce  $\alpha\approx 1$ will be successful at spreading between cells. Setting

\[
\frac{\beta\lambda}{da}=4n
\]
\\
and replacing this along with $b_{opt}=a/\sqrt{n(n-1)}$ in Eq. 22 we recover

\begin{equation}
\bar{s}=n-\frac{1}{2}+\sqrt{\frac{1}{4}+4n\sqrt{n(n-1)}}
\end{equation}
\\
which, under Taylor expansion in $1/n$, gives

\begin{equation}
\bar{s}=3n-1+O(n^{-1})
\end{equation}
\\
and the optimal prion is around 3 times its minimum length.

\section*{Inter-cellular dynamics}

Having shown that there is a tradeoff between the ability to invade an empty call and intra-cellular competition between prions, we next consider the inter-cellular dynamics between a pair of strains spreading through a well mixed population of cells. We adopt a compartmental model in which the density of uninfected cells is $S$ and the density of cells infected with prion strain 1 is $P_1$ while the density of cells infected with prion strain 2 is $P_2$. 

\subsection*{Compartmental model of prion dynamics}

Under this model we assume that $P_1$ can invade $P_2$, while $P_2$ is better at invading uninfected cells that $P_1$. And so the dynamics of the compartmental model are as described in the main text (Eq 3).
Susceptible cells undergo logistic growth, while they are invaded by strain 1 at rate $P_1$ and strain 2 at rate $B_2$ where $B_1<B_2$. Strain 2 is invaded by strain 2 at rate $D$, and infected cells die at rate $\nu_1$ and $\nu_2$ respectively. Eq. 35 has a steady state solution with coexistence of strains at density

\begin{eqnarray}
\nonumber \bar{S}&=&1-\frac{B_2\nu_1-B_1\nu_2}{D}\\
\nonumber \bar{P}_1&=&\frac{D(B_2-\nu_2)-B_2(B_2\nu_1-B_1\nu_2)}{D^2}\\
\bar{P}_2&=&\frac{B_1(B_2\nu_1-B_1\nu_2)-D(B_1-\nu_1)}{D^2}
\end{eqnarray}
\\
Which produces viable solutions provided $B_2\nu_1-B_1\nu_2>0$ along with $\frac{B_1(B_2\nu_1-B_1\nu_2)}{B_1-\nu_1}>D>\frac{B_2(B_2\nu_1-B_1\nu_2)}{B_2-\nu_2}$ where the second condition always holds provided the first holds and $B_1>\nu_1$ and $B_2>\nu_2$ (i.e each strain is able to spread individually). Alternately in $B_1<\nu_1$, only the lower bound is necessary. The stability of this model can be assessed numerically and stable coexistence is observed between strains (main text Figure 5).

\subsection*{Slow prion model}

Next we make the simplifying assumption that the spread of prions between cells is slow, such that the density of susceptible cells can be assumed to move instantaneously to equilibrium $\bar{S}\approx 1$. And so we recover the 2-D system of equations.

\begin{eqnarray}
\nonumber \frac{dP_1}{dt}&=&B_1P_1+P_IP_2D-\nu_1P_1\\
\frac{dP_2}{dt}&=&B_2P_2-P_IP_2D-\nu_2P_2
\end{eqnarray}
\\
which has coexistent steady state

\begin{eqnarray}
\nonumber \bar{P}_1&=&\frac{B_2-\nu_2}{D}\\
\bar{P}_2&=&\frac{\nu_1-B1}{D}
\end{eqnarray}
\\
which is viable if $B_2>\nu_2$ and $B_1<\nu_1$. The eigenvalues associated with this equilibrium are given by

\begin{equation}
\Lambda=\pm i D\sqrt{\bar{P_1}\bar{P_2}}
\end{equation}
\\
i.e the system is characterized by limit cycles.

\section*{Spatial Model}

\subsection*{Model description and simulations.} To investigate how spatial factors affect coexistence dynamics, we develop a spatially extended version of our ODE model. We employ a square lattice with periodic boundary conditions to host an agent-based simulation whose rules are defined by the processes described in our ODE model. Each lattice site represents a cell which can either be infected or susceptible; i.e. each cell can either be empty, susceptible, infected by strain 1 ($P_{1}$), or infected by strain 2 ($P_{2}$). Interactions among each cell can be defined by the following processes:

\begin{center}
\begin{parbox}[t]{.3\linewidth}{\textsf{\underline{Summary of SNPM Reactions} \\
(1) $S$ + $\emptyset$ $\xrightarrow{a}$ $S + S$ \\
(2) $S + P_{1}$ $\xrightarrow{\beta_{01}}$ $P_{1} + P_{1}$\\
(3) $S + P_{2}$ $\xrightarrow{\beta_{02}}$ $P_{2} + P_{2}$\\
(4) $P_{1} + P_{2}$ $\xrightarrow{\beta_{12}}$ $P_{2} + P_{2}$ \\
(5) $P_{1} + P_{2}$ $\xrightarrow{\beta_{21}}$ $P_{1} + P_{1}$\\
(6) $P_{1}$ $\xrightarrow{\nu_{1}}$ $\emptyset$ \\
(7) $P_{2}$ $\xrightarrow{\nu_{2}}$ $\emptyset$ \\
\hfill
}}
\\
\end{parbox}
\end{center}

Where $\emptyset$ denotes an empty site on the lattice and all rates consider the processes described in the ODE model. Thus, susceptible cells can reproduce with rate $a$ to an empty neighbor cell, infection can occur with a corresponding rate $\beta$ and death of an infected cell can occur with the appropriate rate $\nu$. One timestep is completed when on average all cells have participated in an event. Each Monte Carlo simulation is defined by the following stochastic algorithm:

\begin{enumerate}
  \item Select a cell on the lattice at random and transition into one of the following events with uniform probability:
  \item \textbf{Birth of a susceptible cell}. If the cell is susceptible to infection, randomly select one of the eight adjacent sites; if that site is empty, then with probability $a$ a new susceptible cell is placed on the neighboring site. 
  \item \textbf{Infection of susceptible with strain 1}. If the cell is infected with strain 1, choose a neighboring site at random; if that neighbor is susceptible to infection, then with probability $\beta_{01}$ it will become infected with strain 1.
  \item \textbf{Infection of susceptible with strain 2}. If the cell is infected with strain 2, choose a neighboring site at random; if that neighbor is susceptible to infection, then with probability $\beta_{02}$ it will become infected with strain 2.
  \item \textbf{Invasion of resident by strain 1}. If the cell is infected with strain 1, choose a neighboring site at random; if that neighbor is infected with strain 2, then with probability $\beta_{21}$ it will become infected with strain 1.
  \item \textbf{Invasion of resident by strain 2}. If the cell is infected with strain 2, choose a neighboring site at random; if that neighbor is infected with strain 1, then with probability $\beta_{12}$ it will become infected with strain 2.
  \item \textbf{Death of cell infected with strain 1}. If the cell is infected with strain 1, then with probability $\nu_{1}$ the site becomes empty.
  \item \textbf{Death of cell infected with strain 2}. If the cell is infected with strain 2, then with probability $\nu_{2}$ the site becomes empty.
\end{enumerate}

We observe (Figure S1)similar dynamics to the ODE model in our spatial extension. The snapshots at various time steps show the early spreading dynamics of both strains: initially, the transmissible strain (strain 1, orange) starts to infect susceptible cells (blue). The virulent strain (strain 2, red) then starts to invade cells infected with the transmissible strain. Cells die due to the high virulence of strain 2 and the low frequencies of sites infected with the transmissible strain. As a result, susceptible cells start to colonize the available empty sites. Such a cycle causes the initial dampening oscillations, until the virulent strain starts to cluster around sites saturated with susceptible cells and the system reaches an equilibrium. We can observe this behavior in the phase portrait in (d), which shows initial oscillations until finally dampening and reaching equilibrium.

\begin{figure}
\centering
\includegraphics[scale=0.6]{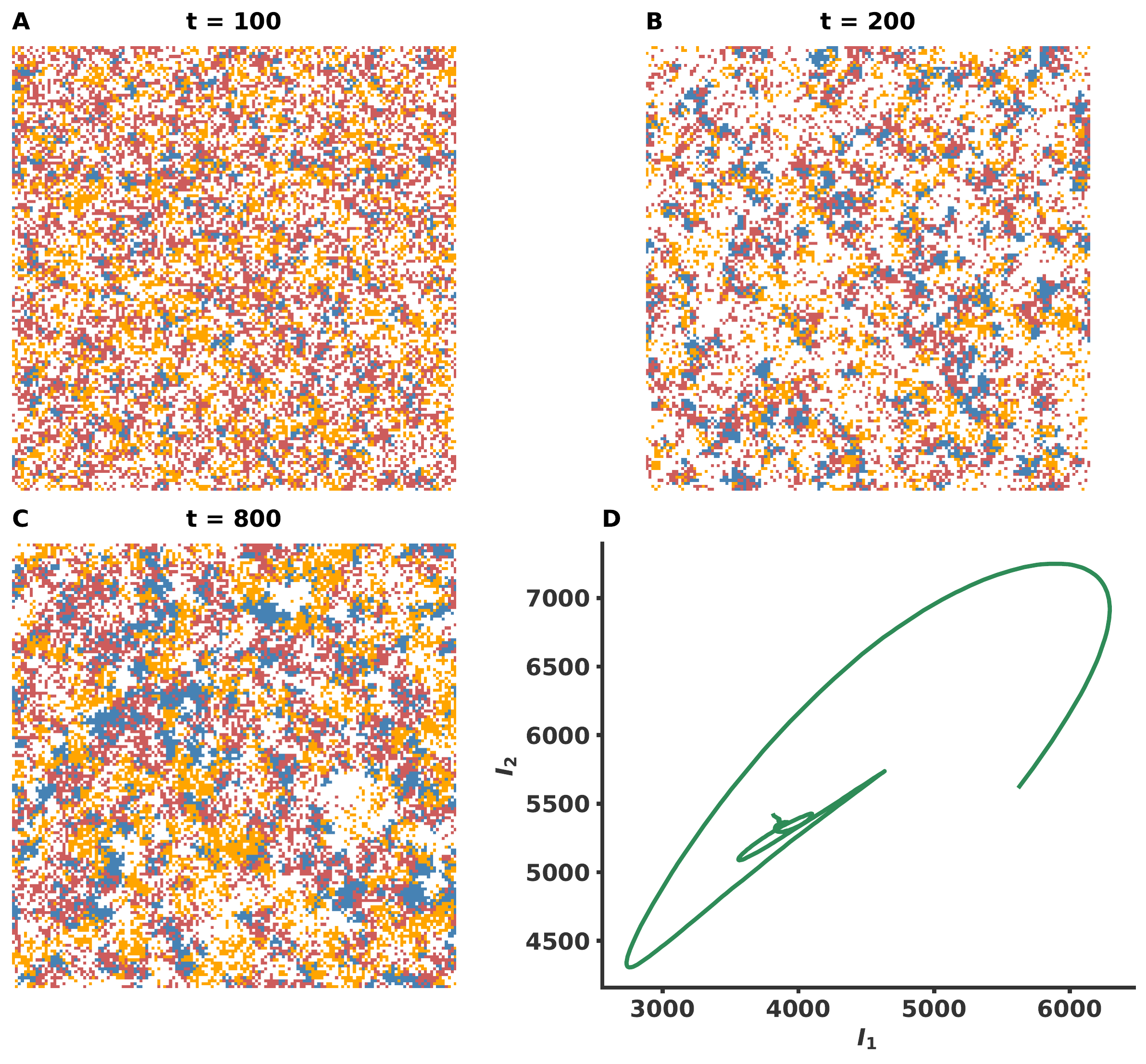}
\caption*{\small Figure S1: \textbf{Spatial model simulation.} (a), (b) and (c) show snapshots of the spatial distribution of cells at different time steps for a single simulation run of the spatial model on a 150x150 square lattice with periodic boundary conditions. (d) shows the phase portrait for both strains after 2000 time steps, which was derived from the average counts of an ensemble of 1000 simulations. 
The event probabilities for these simulations are: $\beta_{01}$ = 0.8, $\beta_{02}$ = 0.5, $\beta_{12}$ = 0.8, $\beta_{21}$ = 0.5, $\nu_{1}$ = 0.05, $\nu_{2}$ = 0.1, $a$ = 0.95. Susceptible cells are shown in blue, those infected with strain 1 in orange, those infected with strain 2 in red and empty sites are indicated in white.}
\end{figure}


\begin{thebibliography}{10}
\expandafter\ifx\csname url\endcsname\relax
  \def\url#1{\texttt{#1}}\fi
\expandafter\ifx\csname urlprefix\endcsname\relax\def\urlprefix{URL }\fi
\providecommand{\bibinfo}[2]{#2}
\providecommand{\eprint}[2][]{\url{#2}}

\bibitem{scheckel2018prions}
\bibinfo{author}{Scheckel, C.} \& \bibinfo{author}{Aguzzi, A.}
\newblock \bibinfo{title}{Prions, prionoids and protein misfolding disorders}.
\newblock \emph{\bibinfo{journal}{Nature Reviews Genetics}}
  \textbf{\bibinfo{volume}{19}}, \bibinfo{pages}{405--418}
  (\bibinfo{year}{2018}).

\bibitem{nowak1998prion}
\bibinfo{author}{Nowak, M.~A.}, \bibinfo{author}{Krakauer, D.~C.},
  \bibinfo{author}{Klug, A.} \& \bibinfo{author}{May, R.~M.}
\newblock \bibinfo{title}{Prion infection dynamics}.
\newblock \emph{\bibinfo{journal}{Integrative Biology: Issues, News, and
  Reviews: Published in Association with The Society for Integrative and
  Comparative Biology}} \textbf{\bibinfo{volume}{1}}, \bibinfo{pages}{3--15}
  (\bibinfo{year}{1998}).

\bibitem{masel1999quantifying}
\bibinfo{author}{Masel, J.}, \bibinfo{author}{Jansen, V.~A.} \&
  \bibinfo{author}{Nowak, M.~A.}
\newblock \bibinfo{title}{Quantifying the kinetic parameters of prion
  replication}.
\newblock \emph{\bibinfo{journal}{Biophysical chemistry}}
  \textbf{\bibinfo{volume}{77}}, \bibinfo{pages}{139--152}
  (\bibinfo{year}{1999}).

\bibitem{masel2001measured}
\bibinfo{author}{Masel, J.} \& \bibinfo{author}{Jansen, V.~A.}
\newblock \bibinfo{title}{The measured level of prion infectivity varies in a
  predictable way according to the aggregation state of the infectious agent}.
\newblock \emph{\bibinfo{journal}{Biochimica et Biophysica Acta (BBA)-Molecular
  Basis of Disease}} \textbf{\bibinfo{volume}{1535}}, \bibinfo{pages}{164--173}
  (\bibinfo{year}{2001}).

\bibitem{payne1998spatial}
\bibinfo{author}{Payne, R.~J.} \& \bibinfo{author}{Krakauer, D.~C.}
\newblock \bibinfo{title}{The spatial dynamics of prion disease}.
\newblock \emph{\bibinfo{journal}{Proceedings of the Royal Society of London.
  Series B: Biological Sciences}} \textbf{\bibinfo{volume}{265}},
  \bibinfo{pages}{2341--2346} (\bibinfo{year}{1998}).

\bibitem{tanaka2006physical}
\bibinfo{author}{Tanaka, M.}, \bibinfo{author}{Collins, S.~R.},
  \bibinfo{author}{Toyama, B.~H.} \& \bibinfo{author}{Weissman, J.~S.}
\newblock \bibinfo{title}{The physical basis of how prion conformations
  determine strain phenotypes}.
\newblock \emph{\bibinfo{journal}{Nature}} \textbf{\bibinfo{volume}{442}},
  \bibinfo{pages}{585--589} (\bibinfo{year}{2006}).

\bibitem{lemarre2019generalizing}
\bibinfo{author}{Lemarre, P.}, \bibinfo{author}{Pujo-Menjouet, L.} \&
  \bibinfo{author}{Sindi, S.~S.}
\newblock \bibinfo{title}{Generalizing a mathematical model of prion
  aggregation allows strain coexistence and co-stability by including a novel
  misfolded species}.
\newblock \emph{\bibinfo{journal}{Journal of mathematical biology}}
  \textbf{\bibinfo{volume}{78}}, \bibinfo{pages}{465--495}
  (\bibinfo{year}{2019}).

\bibitem{li2010darwinian}
\bibinfo{author}{Li, J.}, \bibinfo{author}{Browning, S.},
  \bibinfo{author}{Mahal, S.~P.}, \bibinfo{author}{Oelschlegel, A.~M.} \&
  \bibinfo{author}{Weissmann, C.}
\newblock \bibinfo{title}{Darwinian evolution of prions in cell culture}.
\newblock \emph{\bibinfo{journal}{Science}} \textbf{\bibinfo{volume}{327}},
  \bibinfo{pages}{869--872} (\bibinfo{year}{2010}).

\bibitem{krishnan2005structural}
\bibinfo{author}{Krishnan, R.} \& \bibinfo{author}{Lindquist, S.~L.}
\newblock \bibinfo{title}{Structural insights into a yeast prion illuminate
  nucleation and strain diversity}.
\newblock \emph{\bibinfo{journal}{Nature}} \textbf{\bibinfo{volume}{435}},
  \bibinfo{pages}{765--772} (\bibinfo{year}{2005}).

\bibitem{collinge2007general}
\bibinfo{author}{Collinge, J.} \& \bibinfo{author}{Clarke, A.~R.}
\newblock \bibinfo{title}{A general model of prion strains and their
  pathogenicity}.
\newblock \emph{\bibinfo{journal}{Science}} \textbf{\bibinfo{volume}{318}},
  \bibinfo{pages}{930--936} (\bibinfo{year}{2007}).

\bibitem{collinge2010prion}
\bibinfo{author}{Collinge, J.}
\newblock \bibinfo{title}{Prion strain mutation and selection}.
\newblock \emph{\bibinfo{journal}{Science}} \textbf{\bibinfo{volume}{328}},
  \bibinfo{pages}{1111--1112} (\bibinfo{year}{2010}).

\bibitem{Mullon:2016aa}
\bibinfo{author}{Mullon, C.}, \bibinfo{author}{Keller, L.} \&
  \bibinfo{author}{Lehmann, L.}
\newblock \bibinfo{title}{Evolutionary stability of jointly evolving traits in
  subdivided populations}.
\newblock \emph{\bibinfo{journal}{Am Nat}} \textbf{\bibinfo{volume}{188}},
  \bibinfo{pages}{175--95} (\bibinfo{year}{2016}).

\bibitem{bonhoeffer1996curse}
\bibinfo{author}{Bonhoeffer, S.}, \bibinfo{author}{Lenski, R.~E.} \&
  \bibinfo{author}{Ebert, D.}
\newblock \bibinfo{title}{The curse of the pharaoh: the evolution of virulence
  in pathogens with long living propagules}.
\newblock \emph{\bibinfo{journal}{Proceedings of the Royal Society of London.
  Series B: Biological Sciences}} \textbf{\bibinfo{volume}{263}},
  \bibinfo{pages}{715--721} (\bibinfo{year}{1996}).

\bibitem{bonhoeffer1994mutation}
\bibinfo{author}{Bonhoeffer, S.} \& \bibinfo{author}{Nowak, M.~A.}
\newblock \bibinfo{title}{Mutation and the evolution of virulence}.
\newblock \emph{\bibinfo{journal}{Proceedings of the Royal Society of London.
  Series B: Biological Sciences}} \textbf{\bibinfo{volume}{258}},
  \bibinfo{pages}{133--140} (\bibinfo{year}{1994}).

\bibitem{prusiner1998prions}
\bibinfo{author}{Prusiner, S.~B.}
\newblock \bibinfo{title}{Prions}.
\newblock \emph{\bibinfo{journal}{Proceedings of the National Academy of
  Sciences}} \textbf{\bibinfo{volume}{95}}, \bibinfo{pages}{13363--13383}
  (\bibinfo{year}{1998}).

\bibitem{rutherford1998hsp90}
\bibinfo{author}{Rutherford, S.~L.} \& \bibinfo{author}{Lindquist, S.}
\newblock \bibinfo{title}{Hsp90 as a capacitor for morphological evolution}.
\newblock \emph{\bibinfo{journal}{Nature}} \textbf{\bibinfo{volume}{396}},
  \bibinfo{pages}{336--342} (\bibinfo{year}{1998}).

\bibitem{masel2013q}
\bibinfo{author}{Masel, J.}
\newblock \bibinfo{title}{Q\&a: evolutionary capacitance}.
\newblock \emph{\bibinfo{journal}{BMC biology}} \textbf{\bibinfo{volume}{11}},
  \bibinfo{pages}{1--4} (\bibinfo{year}{2013}).

\bibitem{nelson2018evolutionary}
\bibinfo{author}{Nelson, P.} \& \bibinfo{author}{Masel, J.}
\newblock \bibinfo{title}{Evolutionary capacitance emerges spontaneously during
  adaptation to environmental changes}.
\newblock \emph{\bibinfo{journal}{Cell reports}} \textbf{\bibinfo{volume}{25}},
  \bibinfo{pages}{249--258} (\bibinfo{year}{2018}).

\bibitem{aguzzi2009transcellular}
\bibinfo{author}{Aguzzi, A.} \& \bibinfo{author}{Rajendran, L.}
\newblock \bibinfo{title}{The transcellular spread of cytosolic amyloids,
  prions, and prionoids}.
\newblock \emph{\bibinfo{journal}{Neuron}} \textbf{\bibinfo{volume}{64}},
  \bibinfo{pages}{783--790} (\bibinfo{year}{2009}).

\bibitem{aguzzi2016cell}
\bibinfo{author}{Aguzzi, A.} \& \bibinfo{author}{Lakkaraju, A.~K.}
\newblock \bibinfo{title}{Cell biology of prions and prionoids: a status
  report}.
\newblock \emph{\bibinfo{journal}{Trends in cell biology}}
  \textbf{\bibinfo{volume}{26}}, \bibinfo{pages}{40--51}
  (\bibinfo{year}{2016}).

\bibitem{maji2009functional}
\bibinfo{author}{Maji, S.~K.} \emph{et~al.}
\newblock \bibinfo{title}{Functional amyloids as natural storage of peptide
  hormones in pituitary secretory granules}.
\newblock \emph{\bibinfo{journal}{Science}} \textbf{\bibinfo{volume}{325}},
  \bibinfo{pages}{328--332} (\bibinfo{year}{2009}).

\bibitem{otzen2019functional}
\bibinfo{author}{Otzen, D.} \& \bibinfo{author}{Riek, R.}
\newblock \bibinfo{title}{Functional amyloids}.
\newblock \emph{\bibinfo{journal}{Cold Spring Harbor perspectives in biology}}
  \textbf{\bibinfo{volume}{11}}, \bibinfo{pages}{a033860}
  (\bibinfo{year}{2019}).

\bibitem{finder2007amyloid}
\bibinfo{author}{Finder, V.~H.} \& \bibinfo{author}{Glockshuber, R.}
\newblock \bibinfo{title}{Amyloid-$\beta$ aggregation}.
\newblock \emph{\bibinfo{journal}{Neurodegenerative Diseases}}
  \textbf{\bibinfo{volume}{4}}, \bibinfo{pages}{13--27} (\bibinfo{year}{2007}).

\bibitem{roychaudhuri2009amyloid}
\bibinfo{author}{Roychaudhuri, R.}, \bibinfo{author}{Yang, M.},
  \bibinfo{author}{Hoshi, M.~M.} \& \bibinfo{author}{Teplow, D.~B.}
\newblock \bibinfo{title}{Amyloid $\beta$-protein assembly and alzheimer
  disease}.
\newblock \emph{\bibinfo{journal}{Journal of Biological Chemistry}}
  \textbf{\bibinfo{volume}{284}}, \bibinfo{pages}{4749--4753}
  (\bibinfo{year}{2009}).

\bibitem{busche2020synergy}
\bibinfo{author}{Busche, M.~A.} \& \bibinfo{author}{Hyman, B.~T.}
\newblock \bibinfo{title}{Synergy between amyloid-$\beta$ and tau in
  alzheimer’s disease}.
\newblock \emph{\bibinfo{journal}{Nature neuroscience}}
  \textbf{\bibinfo{volume}{23}}, \bibinfo{pages}{1183--1193}
  (\bibinfo{year}{2020}).

\bibitem{maury2009self}
\bibinfo{author}{Maury, C. P.~J.}
\newblock \bibinfo{title}{Self-propagating $\beta$-sheet polypeptide structures
  as prebiotic informational molecular entities: the amyloid world}.
\newblock \emph{\bibinfo{journal}{Origins of Life and Evolution of Biospheres}}
  \textbf{\bibinfo{volume}{39}}, \bibinfo{pages}{141--150}
  (\bibinfo{year}{2009}).

\bibitem{maury2015origin}
\bibinfo{author}{Maury, C. P.~J.}
\newblock \bibinfo{title}{Origin of life. primordial genetics: Information
  transfer in a pre-rna world based on self-replicating beta-sheet amyloid
  conformers}.
\newblock \emph{\bibinfo{journal}{Journal of Theoretical Biology}}
  \textbf{\bibinfo{volume}{382}}, \bibinfo{pages}{292--297}
  (\bibinfo{year}{2015}).

\bibitem{maury2018amyloid}
\bibinfo{author}{Maury, C. P.~J.}
\newblock \bibinfo{title}{Amyloid and the origin of life: self-replicating
  catalytic amyloids as prebiotic informational and protometabolic entities}.
\newblock \emph{\bibinfo{journal}{Cellular and Molecular Life Sciences}}
  \textbf{\bibinfo{volume}{75}}, \bibinfo{pages}{1499--1507}
  (\bibinfo{year}{2018}).

\bibitem{bernhardt2012rna}
\bibinfo{author}{Bernhardt, H.~S.}
\newblock \bibinfo{title}{The rna world hypothesis: the worst theory of the
  early evolution of life (except for all the others) a}.
\newblock \emph{\bibinfo{journal}{Biology direct}}
  \textbf{\bibinfo{volume}{7}}, \bibinfo{pages}{1--10} (\bibinfo{year}{2012}).

\bibitem{jung2003prions}
\bibinfo{author}{Jung, M.}, \bibinfo{author}{Pistolesi, D.} \&
  \bibinfo{author}{Pana, A.}
\newblock \bibinfo{title}{Prions, prion diseases and decontamination.}
\newblock \emph{\bibinfo{journal}{Igiene e sanita pubblica}}
  \textbf{\bibinfo{volume}{59}}, \bibinfo{pages}{331--344}
  (\bibinfo{year}{2003}).
  
  \bibitem{ledur}
\bibinfo{author}{Le~Dur, A.} \emph{et~al.}
\newblock \bibinfo{title}{Divergent prion strain evolution driven by prpc
  expression level in transgenic mice}.
\newblock \emph{\bibinfo{journal}{Nature Communications}}
  \textbf{\bibinfo{volume}{8}}, \bibinfo{pages}{14170} (\bibinfo{year}{2017}).

\end{thebibliography}
\end{document}